\documentclass[sigconf]{acmart}

\usepackage{booktabs}
\usepackage{arydshln}
\usepackage{multirow}
\usepackage{enumitem}

\graphicspath{ {./img/} }
\AtBeginDocument{%
  }

\copyrightyear{2025}
\acmYear{2025}
\setcopyright{rightsretained}
\acmConference[RecSys '25]{Proceedings of the Nineteenth ACM Conference on Recommender Systems}{September 22--26, 2025}{Prague, Czech Republic}
\acmBooktitle{Proceedings of the Nineteenth ACM Conference on Recommender Systems (RecSys '25), September 22--26, 2025, Prague, Czech Republic}\acmDOI{10.1145/3705328.3759300}
\acmISBN{979-8-4007-1364-4/2025/09}

\begin{document}

\title{Semantic IDs for Joint Generative Search and Recommendation}

\author{Gustavo Penha}
\email{gustavop@spotify.com}
\affiliation{%
  \institution{Spotify}
  \city{Delft}
  \country{Netherlands}
}

\author{Edoardo D'Amico}
\email{edoardod@spotify.com}
\affiliation{%
  \institution{Spotify}
  \city{Madrid}
  \country{Spain}
}

\author{Marco De Nadai}
\email{mdenadai@spotify.com}
\affiliation{%
  \institution{Spotify}
  \city{Copenhagen}
  \country{Denmark}
}

\author{Enrico Palumbo}
\email{enricop@spotify.com}
\affiliation{%
  \institution{Spotify}
  \city{Turin}
  \country{Italy}
}

\author{Alexandre Tamborrino}
\email{alexandret@spotify.com}
\affiliation{%
  \institution{Spotify}
  \city{Paris}
  \country{France}
}

\author{Ali Vardasbi}
\email{aliv@spotify.com}
\affiliation{%
  \institution{Spotify}
  \city{Amsterdam}
  \country{Netherlands}
}

\author{Max Lefarov}
\email{mlefarov@spotify.com}
\affiliation{%
  \institution{Spotify}
  \city{Munich}
  \country{Germany}
}

\author{Shawn Lin}
\email{weihsiangl@spotify.com}
\affiliation{%
  \institution{Spotify}
  \city{New York}
  \country{United States}
}

\author{Timothy Heath}
\email{theath@spotify.com}
\affiliation{%
  \institution{Spotify}
  \city{New York}
  \country{United States}
}

\author{Francesco Fabbri}
\email{francescof@spotify.com}
\affiliation{%
  \institution{Spotify}
  \city{Barcelona}
  \country{Spain}
}

\author{Hugues Bouchard}
\email{hb@spotify.com}
\affiliation{%
  \institution{Spotify}
  \city{Barcelona}
  \country{Spain}
}

\renewcommand{\shortauthors}{Penha et al.}

\begin{abstract}
Generative models powered by Large Language Models (LLMs) are emerging as a unified solution for powering both recommendation and search tasks. A key design choice in these models is how to represent items, traditionally through unique identifiers (IDs) and more recently with Semantic IDs composed of discrete codes, obtained from embeddings. While task-specific embedding models can improve performance for individual tasks, they may not generalize well in a joint setting. In this paper, we explore how to construct Semantic IDs that perform well both in search and recommendation when using a unified model. We compare a range of strategies to construct Semantic IDs, looking into task-specific and cross-tasks approaches, and also whether each task should have its own semantic ID tokens in a joint search and recommendation generative model. Our results show that using a bi-encoder model fine-tuned on both search and recommendation tasks to obtain item embeddings, followed by the construction of a unified Semantic ID space provides an effective trade-off, enabling strong performance in both tasks. We hope these findings spark follow-up work on generalisable, semantically grounded ID schemes and inform the next wave of unified generative recommender architectures.

\end{abstract}

\maketitle

\section{Introduction}

\begin{figure}[]
\includegraphics[width=0.4\textwidth]{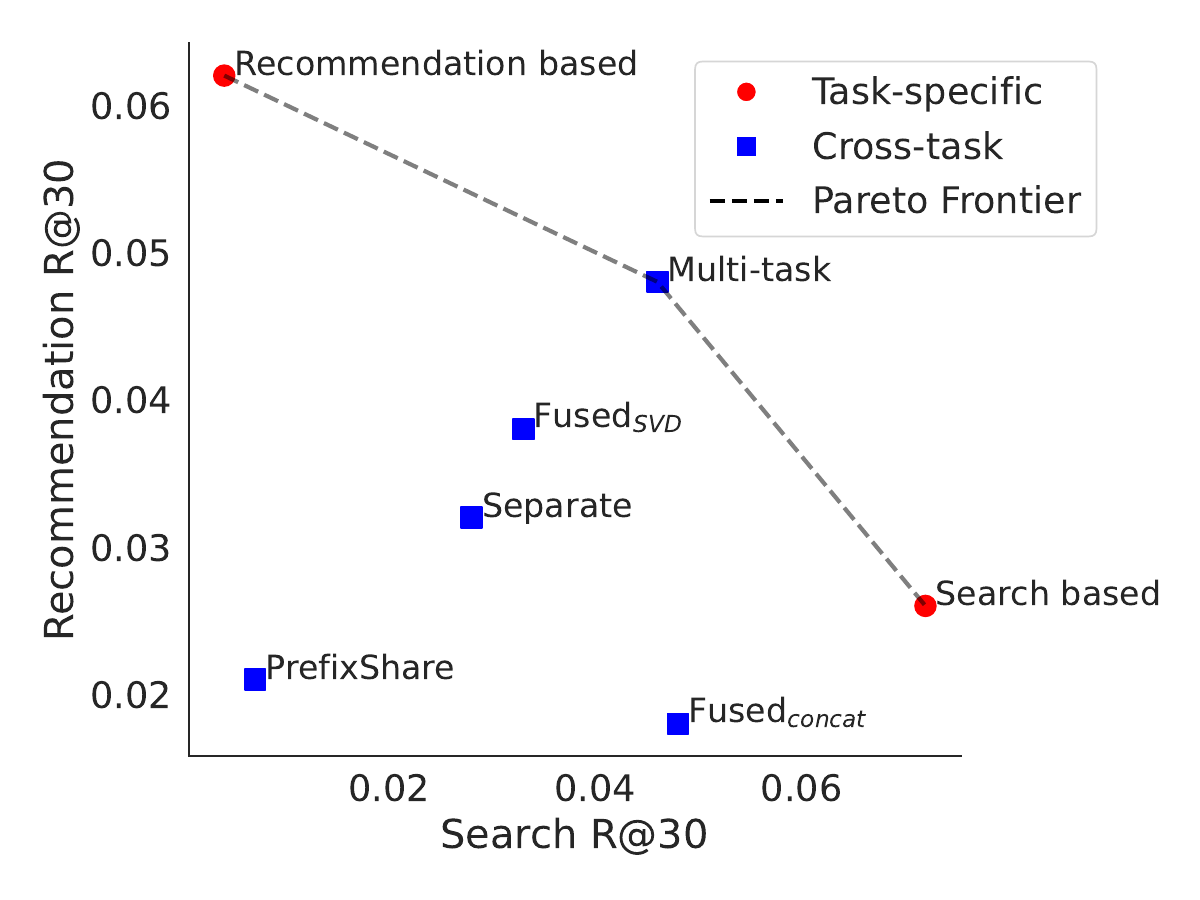}
\caption{Effectiveness of a joint generative model for search an recommendation using different approaches to construct Semantic IDs. We investigate embeddings that take into account both search and recommendation when representing items and constructing Semantic IDs.}
\label{fig:intro}
\end{figure}

Generative models powered by Large Language Models (LLMs) are transforming how we approach both recommendation and search tasks. Rather than building task-specific models, recent work have explored unified generative frameworks that simplify system design and potentially improve generalization across tasks. 

In generative models, item representations must be mapped to discrete tokens that LLMs can consume and produce~\cite{palumbo2025text2tracks, hua2023index,palumbo2025text2tracks,petrov2023generative,doh2025talkplay}. Traditional recommender systems often use unique item identifiers (IDs) that are added to the model's vocabulary, such as in SASRec~\cite{kang2018self}, while others use sequential IDs (e.g. P5~\cite{geng2022recommendation}) based on heuristics, or less token-efficient solutions such as title of the entities involved~\cite{de2020autoregressive}. Most of these approaches require re-training the model whenever new cold start items are added and fall short in industrial settings.

To address this, recent work have proposed using Semantic IDs, sets of discrete tokens generated from pre-trained item embeddings~\cite{rajput2023recommender,tay2022transformer}. These IDs allow items with similar content (embeddings) to share tokens, improving generalization and enabling cold start settings. However, generating effective Semantic IDs critically depends on the embedding space used to construct them. Prior work has shown that fine-tuning embeddings for a specific task, such as recommendation or search, yields the most effective Semantic IDs for that task~\cite{10.1145/3589334.3645477,qu2024tokenreclearningtokenizeid}. 
However, this raises an important question:

\vspace{0.5em}
\noindent\textit{Can we create Semantic IDs that perform well for \textbf{both} search and recommendation in a joint generative model?}
\vspace{0.5em}

This is particularly relevant as a \textbf{joint search and recommedation} (Joint S\&R) approach has emerged as a promising strategy to reduce engineering overhead and improve performance~\cite{bhattacharya2024joint,zheng2024adapting,zamani2018joint,zhao2025unifying,penha2024bridging} by unifying data sources and models that have traditionally been treated as separated silos.

In this paper, we investigate the construction of Semantic IDs in the context of joint search and recommendation generative model. Specifically, we study how different embedding sources, finetuned for search, recommendation or both, impact the downstream performance of the model. We also consider whether it is more beneficial to share or to have task specific tokens to represent each item.

Our findings confirm that there is a fundamental trade-off in constructing Semantic IDs: embeddings finetuned for one task (e.g. recommendation) degrade performance on the other (e.g. search), and vice-versa. This calls into question the growing trend of using task-specific embedding spaces as priors to ID generation in generative models~\cite{10.1145/3589334.3645477,qu2024tokenreclearningtokenizeid}.

We show that a bi-encoder model jointly fine-tuned on both search and recommendation offers a compelling compromise (see Figure~\ref{fig:intro}), yielding Semantic IDs that generalize across tasks with minimal loss in individual task effectiveness. This challenges the conventional wisdom that optimal performance requires the construction of a per-task ID. In contrast, our result suggest that a shared representation space can streamline generative model design without sacrificing quality, especially in multi-task systems.

As LLM-based recommendation continues to evolve, we believe this work offers a timely signal: that better generative performance may come not from task-specific specialization, but from well-constructed, generalizable semantic IDs.

\begin{figure*}[ht!]
\includegraphics[width=0.90\textwidth]{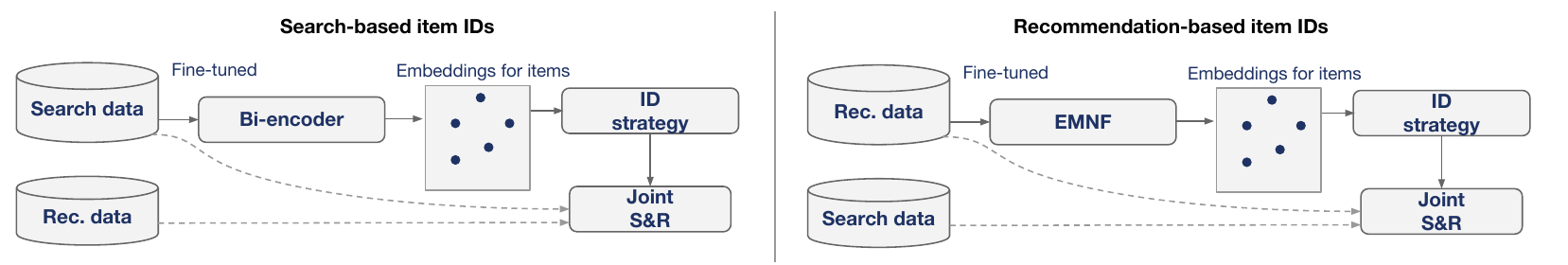}
\caption{Task specific Semantic ID construction methods for a joint generative search and recommendation model. On the left we have Semantic IDs based on a bi-encoder model fine-tuned for search (\textbf{Search based}), whereas on the right we have Semantic IDs based on the embeddings of a collaborative-filtering model fine-tuned for recommendation (\textbf{Rec. based}).}
\label{fig:task_specific}
\end{figure*}

\section{Motivation}

Generative models promise a single architecture for both search and recommendation, yet the previous work optimizes IDs in a task-specific space. We ask: \textit{What happens when those separate optimizations collide in a unified model?}. To answer this, we reproduce the three ID-construction pipelines:

\begin{enumerate}
    \item Content-based: text embeddings such as TIGER~\cite{rajput2023recommender} and DSI~\cite{tay2022transformer}, not fine-tuned for either search or recommendation.
    \item Search-tuned IDs: embeddings fine-tuned for retrieval effectiveness similarly to~\citet{10.1145/3589334.3645477}
    \item Recommendation-tuned IDs: collaborative-filtering model embeddings fine-tuned for recommendation similarly to~\citet{qu2024tokenreclearningtokenizeid}.
\end{enumerate}

\begin{table}[]
\caption{Fine-tuning the embeddings---used for Semantic ID construction---for search and recommendation is effective in a joint generative model. However,  choosing one of the fine-tuned embedding spaces comes at the expense of the other task effectiveness. Bold indicates highest effectiveness while the superscripts denote statistical significance using paired t-tests and Bonferroni correction.}
\label{table:motivation}
\begin{tabular}{@{}rlll@{}}
\toprule
\multicolumn{1}{l}{} & \multicolumn{1}{c}{Embedding space} & \multicolumn{1}{c}{\begin{tabular}[c]{@{}c@{}}Search \\ R@30 ($\pm$ std.)\end{tabular}} & \multicolumn{1}{c}{\begin{tabular}[c]{@{}c@{}}Recommendation\\ R@30 ($\pm$ std.)\end{tabular}} \\ \midrule
\footnotesize{1} & \begin{tabular}[c]{@{}l@{}}Content-based \\ \footnotesize{(e.g. DSI~\cite{tay2022transformer}, TIGER \cite{rajput2023recommender})}\end{tabular} & $0.013\:(\pm0.009)^{}$ & $0.023\:(\pm0.017)^{}$ \\ \hdashline
\footnotesize{2} & \begin{tabular}[c]{@{}l@{}}Search based \\  \footnotesize{(e.g. RIPOR~\cite{10.1145/3589334.3645477})}\end{tabular} & \boldmath$0.072\:(\pm0.028)^{13}$ & $0.026\:(\pm0.017)^{}$ \\ \hdashline
\footnotesize{3} & \begin{tabular}[c]{@{}l@{}}Rec. based \\ \footnotesize{(e.g. TokenRec~\cite{qu2024tokenreclearningtokenizeid})}\end{tabular} & $0.004\:(\pm0.001)^{}$ & \boldmath$0.062\:(\pm0.015)^{12}$ \\ \bottomrule
\end{tabular}
\end{table}

After calculating such embeddings for each item of the catalogue, semantic IDs are obtained by applying an ID strategy that tokenizes each item embedding into a set of discrete tokens. 

Table~\ref{table:motivation} shows the performance of the semantic IDs coming from each embedding space when plugged into our Joint S\&R model. The pattern is clear: \emph{optimizing for one task sacrifices the other}. Search-tuned IDs boost retrieval by $\approx$5x but reduces recommendation performance by 60\%. IDs fine-tuned for recommendation flip the effect. Concurrent work reports the same tension~\cite{shi2025unified}.

These observations set the stage for our contribution: \textit{Can we build a single set of semantic IDs that balances both sides?} In the next section we introduce a bi-encoder jointly fine-tuned on both tasks and demonstrate how its shared embedding space reconciles the trade-off without increasing the computational costs of the Joint S\&R model.

\section{Semantic ID Construction}

In this section we define multiple methods for mapping items to IDs, grouped into (i) \textit{task-specific} and (ii) \textit{cross-task} strategies. 

\subsection{Task-Specific Approaches}
Figure~\ref{fig:task_specific} depicts the pipeline shared by the two task-specific baselines.  Each trains an \textbf{embedding model} on a single supervision signal and then discretizes that embedding with a quantizer model.

\paragraph{Search-based} Following the relevance-based ID construction~\cite{10.1145/3589334.3645477} we train a bi-encoder model on search data $D_S$, composed of query and relevant item pairs using in-batch cross negative samples. We use the concatenated item metadata (e.g. title, description) of the item as its document representation. The resulting embeddings $\mathbf{v}^{\text{search}}$ are then discretized to generate Semantic IDs, which are used for both search and recommendation tasks.

\paragraph{Recommendation-based} Analogously, following TokenRec~\cite{qu2024tokenreclearningtokenizeid}, we train an Efficient Neural Matrix Factorization (ENMF) model~\cite{chen2020efficient} on dataset $D_R$ of interacted items of users to create collaborative-filtering based embeddings $\mathbf{v}^{\text{rec}}$. The embedding of each item is discretized and supplied to the generative model for both tasks.

\subsection{Cross-Task Approaches}
We now describe five methods that \emph{explicitly} incorporate information from \emph{both} tasks.

\begin{figure}[ht!]
\includegraphics[width=0.45\textwidth]{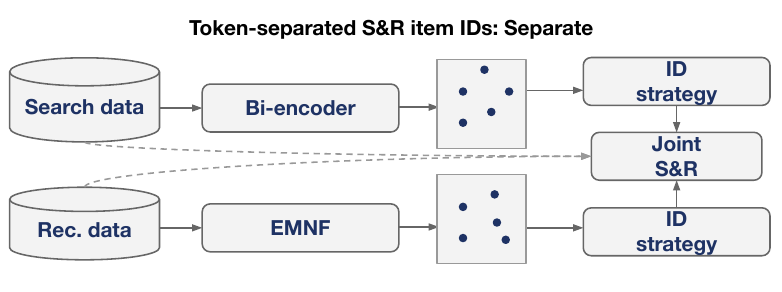}
\caption{Token-separated Semantic IDs approach (\textbf{Separate}), where each task has its own set of semantic IDs constructed from search and recommendation speficic embeddings.} 
\label{fig:separate}
\end{figure}

\paragraph{Token-separated IDs.} Figure~\ref{fig:separate} shows the \textbf{Separate} variant.  We simply prepend task tags to the two task-specific IDs above, yielding tokens: $\text{ID}^{\text{sep}}_i = \langle \mathtt{SEARCH{:}}\text{ID}^{\text{search}}_i,\; \mathtt{REC{:}}\text{ID}^{\text{rec}}_i\rangle .$
This strategy doubles the ID vocabulary size but keeps training simple: search prompts may only output search tokens, while recommendation prompts only recommendation ones.

\paragraph{Prefix-Share IDs.}
Adapting the idea of \citet{shi2025unified}, \textbf{Prefix-share} allocates three codebooks: one shared (\texttt{SHARED}) plus the two task-specific prefixes above.  A single encoder ingests the concatenated embeddings $[\mathbf{v}^{\text{search}}_i;\mathbf{v}^{\text{rec}}_i]$; two decoders learn codebook-specific reconstructions.  The final ID is the concatenation of shared tokens followed by task-specific ones.

\begin{figure*}[]
\includegraphics[width=0.90\textwidth]{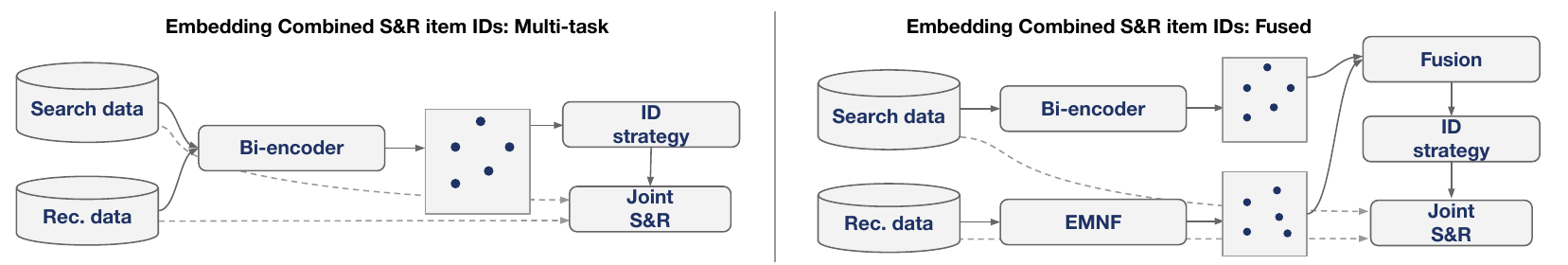}
\caption{Embedding combined approaches where both tasks are considered. On the left we have \textbf{Multi-task} which first trains a Bi-encoder for both search and recommendation and then generates Semantic IDs. On the right we have \textbf{Fused} which combines both embeddings coming from search and recommendation models and then uses the combination to generate Semantic IDs.}
\label{fig:both_tasks}
\end{figure*}

\paragraph{Embedding-Combined IDs.}
Figure~\ref{fig:both_tasks} illustrates three embedding-level fusion strategies.

\begin{itemize}[leftmargin=*]

    \item \textbf{Fused\textsubscript{concat}.}  
          We $\ell_2$-normalize $\mathbf{v}^{\text{search}}_i$ and $\mathbf{v}^{\text{rec}}_i$ and concatenate them:  
          $\mathbf{v}^{\text{concat}}_i = [\mathbf{v}^{\text{search}}_i;\mathbf{v}^{\text{rec}}_i]$.

    \item \textbf{Fused\textsubscript{SVD}.}  
          We again normalize the two embeddings but first reduce the higher-dimensional space with truncated SVD so that both have equal dimensionality $d$.  We then element-wise \emph{add} them:  
          $\mathbf{v}^{\text{svd}}_i = \mathbf{v}^{\text{search}}_i + \mathbf{v}^{\text{rec}}_i$.

    \item \textbf{Multi-task.}  
          We train the bi-encoder on \emph{both} supervision signals: query–item pairs from $D_S$ \emph{and} co-occurring item pairs from $D_R$.  The shared encoder is optimized with the sum of the two contrastive losses, producing embeddings $\mathbf{v}^{\text{mt}}_i$ that carry retrieval \emph{and} collaborative filtering cues.
\end{itemize}

\section{Experimental Setup}

\paragraph{Dataset} Following Penha \emph{et al}~\cite{penha2024bridging} we build a S\&R dataset from MovieLens25M~\cite{harper2015movielens}. The dataset contains 62\,138 movies,  1.24 M user–item interactions (chronologically split; last interaction per user for test) and exactly 20 natural-language queries per item (10 train / 10 test) generated with \texttt{Gemini-2.0-flash}\footnote{We use the following prompt: ``\textit{Your task is to return a list with 10 queries for a given movie (title of the movie, year and description and tags) After generating the initial set of queries, you should also generate a list of the same size with paraphrased of the first queries. The paraphrased queries should be similar to the original queries, but with different words, structure and slight variations in the meaning. The queries should be realistic things that a user would ask to find the movie. The queries should be diverse and cover different aspects of the movie. The queries should not include the title of the movie, but be broader descriptions of the movie and its content. The queries should also contain broad topics, themes and genres of the movie. Movie: \{METADATA}\}''.}. We note that having 10/10 queries for train/test for each item removes the popularity bias of the search dataset. Given that we do not know the true distribution of popularity in search, i.e. there are no real user logs for the MovieLens data, we decided on using the uniform distribution. This means that the search popularity distribution is quite different from the recommendation distribution, and thus we might expect results to be more favorable in real-life distributions with some similarity between popularity distributions. Approaches that use the content of the items (content-based, search based and cross-task ones) use the title, year, description, genres, tags and genome tags~\cite{vig2012tag} to calculate their embeddings. 

\paragraph{Evaluation Metrics}
Given our focus on the retrieval task, we rely on Recall@30. We run every model five times with different random seeds and report the mean recall across different runs. We assess the statistical significance of our results using paired Students' t-tests with a 95\% confidence interval. 

\paragraph{Embedding models} We use the following models to generate item embeddings:
\begin{itemize}[leftmargin=*]
  \item \textbf{Content-based:} we use the pre-trained \texttt{all-mpnet-base-v2} from sentence transformers~\cite{reimers-2019-sentence-bert} on top of the concatenated metadata described in the dataset subsection.
  \item \textbf{For search:} we start from the same pre-trained model, and further fine-tune it on search data with in-batch random negatives (\textit{MultipleNegativesRankingLoss}) using sentence transformers for 5 epochs, batch 512, LR 2e-5, Adam. 
  \item \textbf{For recommendation:} We train ENMF~\cite{chen2020efficient} via RecBole~\cite{recbole[2.0]}, 30 epochs, batch 512, embedding 256, LR 0.001, Adam.
\end{itemize}

\smallskip
\paragraph{Generative model}
\texttt{google/flan-t5-base}~\cite{raffel2020exploring}, trained jointly on S\&R for 3 epochs (LR 0.002, batch 128, AdamW, weight-decay 0.01).  
To increase the number of distinct items retrieved for all generative retrieval models we resort to a diversified beam search approach~\cite{vijayakumar2016diverse}: beam 60, diversity penalty 0.25, 30 groups.

\smallskip
\paragraph{ID tokenisation}
Unless otherwise stated we use two codebooks of size 256 (512 tokens total). Some cross-task methods require additional tokens. \emph{Separate} adds in total 1024 new tokens to the vocabulary as each semantic ID space is treated separatedly. \emph{Prefix-share} has 256 shared tokens and 512 task-specific tokens. 

The tokens are constructed using RQ-KMeans\footnote{\url{https://github.com/facebookresearch/faiss/wiki/Additive-quantizers}} (FAISS residual quantiser~\cite{douze2024faiss}) for all models.  
For ablations we additionally evaluate MiniBatchDictionaryLearning (through Sklearn~\cite{scikit-learn}), ResidualLFQ and RQ-VAE from \texttt{vector-quantize-pytorch}\footnote{\url{https://github.com/lucidrains/vector-quantize-pytorch}}.

\section{Results}

\paragraph{\textbf{Semantic IDs for Joint Search and Recommendation}}
\begin{table}[]
\caption{R@30 of the joint generative model using different Semantic ID construction methods that consider both objectives (search and recommendation). Head indicates the effectiveness for the top 1\% most popular items in the train set, where Torso is the remaining set of items. Search data does not have a popularity bias, i.e. all items have the same number of queries. Bold indicates highest scores, while underline indicates second highest.}
\label{table:main_results}
\begin{tabular}{@{}clllll@{}}
\toprule
\multicolumn{1}{l}{} &  & \multicolumn{1}{c}{Search} & \multicolumn{3}{c}{Recommendation} \\ \midrule
\multicolumn{2}{c}{Semantic ID construction} & \multicolumn{1}{c}{All} & \multicolumn{1}{c}{All} & \multicolumn{1}{c}{Head} & \multicolumn{1}{c}{Torso} \\ \midrule
\multirow{2}{*}{Task-specific} & Search based & \textbf{0.072} & 0.026 & 0.090 & \textbf{0.070} \\
 & Rec. based & 0.004 & \textbf{0.062} & \textbf{0.170} & 0.035 \\ \midrule
\multirow{5}{*}{Cross-task} & Separate & 0.028 & 0.032 & 0.120 & 0.051 \\ 
 & Prefix-share & 0.007 & 0.021 & 0.058 & 0.010 \\ \cmidrule(l){2-6} 
 & Fused$_{concat}$ & \underline{0.048} & 0.018 & 0.045 & 0.041 \\
 & Fused$_{SVD}$ & 0.033 & 0.038 & 0.105 & \underline{0.060} \\ 
 & Multi-task & 0.046 & \underline{0.049} & \underline{0.135} & 0.024 \\
 \bottomrule
\end{tabular}
\end{table}
Table~\ref{table:main_results} displays our results for different Semantic ID construction methods, for search and recommendation.

\paragraph{Task-Specific Approaches.}  The first two rows show the top performers \emph{per task}, as anticipated in the motivation. However, they are based on specialized IDs and cannot provide a satisfactory trade-off for a unified S\&R generative model, as each semantic ID is overfitted to the target task. Interestingly, we see that for the \emph{Torso} entities, the search based embeddings are effective, demonstrating that the popularity of each entity plays an important role in the recommendation data and that for less popular items leaning more on content is effective.

\paragraph{Cross-Task Approaches.} The remaining rows are methods that use embeddings from both tasks. \textbf{Separate} leverages distinct tokens for each task, adding more tokens to the vocabulary of the generative model (task-specific Semantic ID tokens). This means that the knowledge learned from one task cannot be used for the task-specific tokens of the other task, negating the regularization effect in item representations discussed by~\citet{penha2024bridging}. 

\textbf{Prefix-share} also underperforms the other methods, due to the underlying quantization approach not performing well here (see next section on the tokenization method ablation). We observe that both methods under perform methods that first combine the embeddings and then construct Semantic IDs (last three rows).

The results for the fusion methods show that the concatenation of the embeddings can be problematic if one of the embedding spaces is larger than the other (the Bi-encoder model is 386 dimensional while the ENMF has 256 dimensions). The embedding space with larger dimensionality might become more represented. By making the dimensionality the same with \textbf{Fused$_{SVD}$} we improve the effectiveness of the model for recommendation, while degrading the search effectiveness compared to \textbf{Fused$_{concat}$}\footnote{Another solution to have embedding spaces with equal size that we did not explore is to train models with a Matryoshka objective~\cite{kusupati2022matryoshka}, and using the first dimensions only.}. 

Finally, we see that training the encoding model for both search and recommendation tasks (\textbf{Multi-task}) offers an effective trade-off of the search and recommendation effectiveness. A similar solution to inject collaborative filtering embeddings into a content-based model has been proposed by~\citet{vanvcura2024beeformer}. 

We believe this is a promising direction to obtain item representations that work well across search and recommendation problems.

\paragraph{\textbf{Tokenization Method Ablation}}
\label{subsec:ablation}
Table~\ref{table:ablation_tokenization} shows the results for the ablation of the tokenization methods that receive as input embeddings and outputs discrete tokens, while keeping the embedding space fixed coming from \emph{Multi-task} (similar results were found when performing the same ablation with the other approaches for Semantic ID construction). We see that RQ-KMeans is the best method to construct IDs for our dataset, outperforming common approaches such as RQ-VAE. \citet{hong2025eager} also found RQ-VAE unstable and opted for hierarchical k-means in their experiments. We leave for future work the investigation of the reasons why it outperforms learned auto-encoder approaches in this experimental setting. 

\begin{table}[]
\caption{R@30 of the joint generative search and recommendation model for different tokenization methods using embeddings from \textbf{Multi-task} approach (similar results were found when performing the same ablation with the other approaches for Semantic ID construction).}
\label{table:ablation_tokenization}
\begin{tabular}{@{}lll@{}}
\toprule
Method & Search & Rec. \\ \midrule
RQ-KMeans & \textbf{0.046} & \textbf{0.049} \\ \hdashline
Dictionary encoding & 0.019 & 0.029 \\
ResidualLFQ & 0.018 & 0.023 \\
RQ-VAE & 0.002 & 0.024 \\ \bottomrule
\end{tabular}
\end{table}

\section{Conclusion}

In this paper we show that the way Semantic IDs are built is a decisive factor for the effectiveness of a unified generative model that serves both search and recommendation tasks, systematically comparing \emph{task-specific}, \emph{token-separated}, and \emph{embedding-combined} constructions.

We find that task-specific Semantic IDs excel only in isolation, whereas the \emph{cross-task} ones deliver a balanced, high-quality solution without inflating the token budget. Ablation of discretisation methods further shows that a lightweight RQ-KMeans tokeniser outperforms VQ-VAE variants.

These observations provide \emph{early empirical evidence} that unifying item representations is not only feasible but advantageous. This insight is critical as LLM-based retrieval and recommendation systems are converging in practice~\cite{bhattacharya2024joint}.  
By highlighting a practical route toward shared Semantic IDs, our study offers a timely snapshot of where the field is heading and points to several open questions on representation learning, token efficiency and cold-start robustness.

We hope these findings spark follow-up work on generalisable, semantically grounded ID schemes and inform the next wave of unified generative recommender architectures.

\bibliographystyle{ACM-Reference-Format}
\bibliography{sample-base}

\end{document}